# ETHICAL CONSIDERATIONS ON NANOTECHNOLOGY

## Manuel Alberto M. Ferreira and José António Filipe

Instituto Universitário de Lisboa (ISCTE-IUL),
Business Research Unit (BRU-IUL) and Information Sciences,
Technologies and Architecture Research Center (ISTAR-IUL),
Lisboa, Portugal

## ABSTRACT

Since a significant time ago, although time runs very fast, nanotechnology transformed from one of the most promising scientific hopes in uncountable human domains into a marvelous certainty. Innumerable scientific studies in several areas of knowledge were made since nanoscale emergence, carrying their contribution to the nanoscience development, leading to a great development of technical and scientific knowledge but also raising numerous problems in the ethical field. In this chapter, nanotechnology is discussed both in terms of ethics and in terms of borders that nanotechnology applications must satisfy and concluding notes are presented, highlighting the results of the analysis. Significant considerations are made on the close connection between ethics and the nanotechnology and the effects over the society and values.

**Keywords:** nanotechnology, society, ethics

# INTRODUCTION

Bearing in mind that nanotechnology is the skill to work at the atomic, molecular and supramolecular levels (at a scale of approximately 1 – 100 nm), the enormous possibilities raised by the work of nanomaterials can be easily understood. The creation and the use of these material structures, devices and systems with essentially new properties and functions result from their very small structure, with applications in several areas, such as bioprocessing in industries; molecular medicine, considering physical, chemical, biological, and medical techniques used to describe molecular structures and mechanisms, often referred as personalized medicine; analyzing the health effect of nanostructures in the environment; improving food and agricultural systems; in summary: improving human performance at many fields, considering the different branches of science and the applications at many dimensions and scales.

Indeed nanotechnology, for instance, supplies the tools and the technology, allowing that researches lead to the transformation of biological systems, to the extent that biology provides models and bio-assembled components to nanotechnology.

Another relevant example: nanoscales, used in biosystems, contribute to enhance very innovative and promising results in medical area; improvements in telemedicine and on health may be expected with new systems operation and new nanotechniques, much less invasive and therefore much less uncomfortable for patients.

However, ethical as far as legal and social implications are postured. This discussion is as old as the issue itself but the boundaries are too much large, even large enough to find out the difficulties on finding definitive answers to this discussion. In the next section this issue is approached in general terms. Follow a section on ethical human dignity and in the next two sections are approached the themes of nanotechnology and environment, first, and of ethics and nanotecnology in the workplace. Then a framework for ethical analysis in nanotechnology is presented and this work finish with a concluding remarks section. In the end a short bibliography on these subjects is presented.

# ETHICS AND NANOTECNOLOGY

Computers have brought an increase in the ability to type and made handwriting more and more superfluous, even obsolete. Mobiles and their miniaturization brought the ability to generally communicate from anywhere with anybody, making more and more useless the traditional phones. Nanotechnology, because of the negligible dimension of devices this technology proportionate, will spread the use of devices that now are nuisance. For instance, the pacemaker that presently is used like a "last resource", certainly soon will be used in a much larger scale because it is hopeful that will be possible to design it at a nanoscale level.

However ethical questions quickly rise with the development of nanotechnology:

- What are the borders for this kind of progress?
- Until where it is admissible and wise to go?
- What are the implications in a later phase of development – obviously unknown - of these technologies?

Global discussion has begun about this theme since the beginning of the discoveries about this subject. Legal, ethical and social implications are irreversible and the discussion is usually asked to be hard and not pacific, needing urgently great clarifications.

Vanessa Nurock - see citation in Bensaude-Vincent (2010)- questions the standard view of an ethics for nanotechnology. She argues that none of the current trends in the discipline of ethics would qualify for application to nanotechnology. Then considering that neurotechnology – a rapidly growing field in the intersection between nano and biotechnology – can affect moral capacities of the brain, she suggests that ethics itself may be affected by nanotechnology. And she leaves as open the question of a co-construction of ethics and bionanotechnology. In short: new approaches must be considered, but which ones?

In Freitas (2005) it is affirmed that the society should be "able to muster the collective financial and moral courage to allow such extraordinarily powerful medicine to be deployed for human betterment, with due regard to essential ethical considerations".

At EU, there are some important concerns and discussion about the development of the integration of human beings and artificial (software/hardware) entities. A funded project in this area (ETHICBOTS) was created to promote and to coordinate a multidisciplinary group of researchers into artificial intelligence, robotics, anthropology, moral philosophy, philosophy of science, psychology, and cognitive science, with the common purpose of identifying and analyzing techno-ethical issues concerned precisely with the integration of human beings and artificial (software/hardware) entities.

Next generations, in the perspective of nanomaterials utilization, will obtain many benefits in all these issues: health, economic, environmental, etc. By using this kind of advanced technology, with many more applications and much welcoming using, the conditions this technology offers will be very environmental friendly as well as already seen.

Actually, there are irreversible and long-term impact consequences for forthcoming generations and for the environment. However, there are risks, as well. But in truth there are incredible potential advantages. And the discussion is all over the place. That borders must be considered is, of course, evident and there has to be an achieved balance between the path found to the new dimension of knowledge and the contours specific for the problem. The rub of the question is exactly in what measure criticisms may have into account the technology potential.

## Ethical Human Dignity

Despite the high feasibility for the economy and the environment, there are some considerations regarding the ethical, human dignity and moral borders on nanotechnology that should be taken into account.

It is interesting to have a look to the definition of human dignity and its connection to moral conveyed in the "opinion of the European Group on ethics in science and new technologies to the European Commission", 2005, when discussing the ethical aspects of information and communication technologies (ICT) implants in the human body. The group makes allusion to the EU's draft Treaty that Establishes a Constitution for Europe, stating that "*Human dignity is inviolable. It must be respected and protected*" (Article II-61), and goes on to explain that "*the dignity of the human person is not only a fundamental right in itself but constitutes the real basis of fundamental rights*" (Declaration concerning the explanations relating to the Charter of Fundamental Rights). This group declares yet that this explanation does not strictly define human dignity and that many writers have attempted to fill this gap. One such attempt suggests that human dignity is defined as follows: "*the exalted moral status which every being of human origin uniquely possesses. Human dignity is a given reality, intrinsic to human substance, and not contingent upon any functional capacities which vary in degree. (...) The possession of human dignity carries certain immutable moral obligations. These include, concerning the treatment of all other human beings, the duty to preserve life, liberty, and the security of persons, and concerning animals and nature, responsibilities of stewardship.*"

The introduction of new devices to go further above the natural capabilities of the human being brings lots of alarms to scientific community. Desires, challenges but also concerns are always present… What are the consequences? To human beings, to life itself, to other living beings, to environment, to biodiversity, to progress, to the civilization as a whole...

Which are the consequences about the limit for the human being as human? Are humans creating a new being? Are humans prepared to get different, biologically and to become another thing? Are humans creating further "nano-digital devices" between humans that have access to these advanced technologies and humans that have not access? This discussion is recognized but probably the ethical discussion has not been enough promoted. And possibly no one is enough prepared to participate effectively and wisely on it. Much reflection, study and information is still needed.

## Nanotecnology and Environment

In the long term, which are the consequences that nanotechnology has in the environment? Is it possible to predict that less damage in nature can be done if nanomaterials are considered instead of large equipment are used? Besides, once the need of inputs to produce these devices is lesser, probably there is an additional advantage on this perspective for nature preservation. But what is the reverse of the medal?

Let's have a look on the following situation:

- There is a growing consensus today that it is considered priority should be given to the development of alternative energies over the classical energies. It turns out that the classic energies price of production is cheaper than that of the alternative energies. But in the case of solar energy, produced with nano-photovoltaic panels, the price, though higher, is already comparable to, say, hydroelectric energy. In this case, this difference in costs can be compared with the environmental damage caused by the construction of dams. And even with the damage due to the pollution produced by the operation of the fuel oil plants. There is here an ethical problem in which one balances between paying a little more and preserving the environment or paying less and destroying it, in which the interests installed with the classical energies will hardly abdicate their positions without struggle, see (Ferreira et al., 2014).

Really, the implications and directions of nanotechnology need further discussion. The consequences of avoiding the discussion may be severe for society, considering rejection or fear of the effects. So it is pertinent to ask:

- Is there a need for a regulatory authority?
- Which kind of mechanisms are needed to regulate the nanotechnologies developments area?

## ETHICS AND NANOTECNOLOGY IN THE WORKPLACE

A theme urgently needing study and reflection concerns the potentially harmful effects on workers' health due to their exposure to nanoparticles in companies dealing with nanotechnology. Indeed, very little, to say nothing, is known about these harmful effects.

In similar situations the harmful effects on workers' health were only identified after long periods of activity, with the occurrence of real disasters in sanitary terms. One has to learn from history and try to avoid situations such as, for example, atomic energy in which, only relatively late, the harmful effects of radiation have been noticed. And only then were the necessary preventions taken, after the damage to health occurred in the workers of the plants and in the neighboring populations. And today humanity, in general, is not very confident on those preventions. However, atomic energy was initially used for war purposes which helps explain, but by no means excuse, this late reaction. The same is not true for nanotechnology, so it is natural that the attitude towards the problem is different.

Thus, in the absence of scientific knowledge on the potentially negative health effects of workplace exposure to nanoparticles, it is necessary to establish guidelines for decision making regarding risks and controls.

Identifying the ethical issues involved certainly helps with proper decision-making. It is therefore imperative for the decision-makers, particularly the employers, the workers, the investors, and the health authorities, to be aware of these issues. As the primary objective of occupational safety and health is the prevention of accidents and diseases in workers, the situations with the ethical implications that affect those most will be, see (Schulte &Salamanca-Buentello, 2007):

(a) Identification and communication of hazards and risks by scientists, authorities, and employers, both from the perspective of each of these groups and from a global perspective;

(b) Awareness and acceptance of risks by workers;

(c) Adequate choice and strict implementation of controls;

(d) Establishment of medical screening programs;

(e) Investment in toxicological research;

(f) Investment in control research.

Ethical issues involve the independent and impartial determination of risks, implacability, autonomy, justice, preservation of privacy, and promotion of respect for people. Their identification and implementation will make the options for decision-makers more evident. And company-wide discussions on the workplace hazards of nanotechnology can be reinforced by a special emphasis on small businesses and the adoption of a global perspective.

Ethical questions about nanotechnology in the workplace arise from the state of knowledge about the dangers of nanomaterials and the hazards and risks they may pose to workers' health. As this knowledge is still incipient, it is imperative to carry out a provisional assessment of the hazards and risks. Workers can exercise their autonomy only if the processes leading to hazard identification and risk assessment are transparent, comprehensible, and credible. Employers will be in accordance with the principles of autonomy, beneficence, non-maleficence, justice, privacy, and respect for persons, as far as, see (Schulte & Salamanca-Buentello, 2007):

(a) Portray hazards and risks without any omission;

(b) Address risks from a social security perspective;

(c) Communicate and dialogue with workers;

(d) Take the necessary measures to control risks so that they are understandable and accepted by workers to ensure their collaboration in their implementation.

## FRAMEWORK FOR ETHICAL ANALYSIS IN NANOTECHNOLOGY

A methodology for conducting an organized and coherent discussion about ethics in this subject is to subdivide it by the following topics of analysis (see Mnyusiwalla et al., 2003): equity, privacy, security, environment, and metaphysical when discussing the relationship in the human-machine binary. About each one of them some considerations may be made. So, assuming these topics, and trying to go as far as possible in the analysis:

### Equity

The classical dichotomy of developed and developing countries raises a new topic for discussion. Indeed, technology and development are closely correlated. Examples of great worries for developing countries are:

- People poverty reduction,
- Energy problem,
- Water problem,
- Health problem,
- Biodiversity maintenance.

Nanotechnology would help these countries to better conditions of population health and way of life. But it is costly to implement any project in this area and more often there are no financial resources available. Developing countries will be lesser benefited with these advances than developed countries. The main problem happens with the access to the benefits of technology considering the poorest and the richest people inside any country. Who may effectively reach the advances of science and take benefit from them? Anyway, it is true that poor people, whatever they are, may benefit, in some measure, from these advances. It may depend on the will of politicians and on the investments capacity (one way of overcoming countries' lack of investment capacity is their indebtedness which, as many current cases show, is nevertheless a generator of future and sometimes very serious financial problems)of some of these countries, for example. A result would be lower needs for energy and a cleaner energy production as well as other environmental benefits: a safer drug delivery, the improvement on health through for example better prevention, diagnosis and treatment.

## Privacy

At this level, it is to consider the enormous innovations which can be reached, by improving surveillance devices or other devices that restrict human privacy. An individual may be facing situations of "invisible" microphones and cameras. This is a real problem if privacy is intended to be assured. Wireless monitoring provides healthcare providers and patients to be mobile and being able to exchange data when needed. But there is the reverse of the medal: privacy violation? It is true that wireless communication and new physiological sensors bring enormous implications in e-health. One point needed is about the necessary discussion on privacy and security.

## Security

Surely new powerful weapons will be available very soon, as far as "invisible" microphones and cameras will be available. Will this ensure new ways of security or will be added to a potential "arsenal of bio-terrorism and techno-terrorism or even nano-terrorism?" (see Mnyusiwalla et al., 2003). Mnyusiwalla et al., 2003 also ask "who will regulate the direction of research in defensive and offensive military NT [nanotechnology]? How much transparency will be necessary in government and private NT initiatives to avoid misuses? There are also very interesting legal questions involving monitoring, ownership, and control of invisible objects".

## Environment

This topic was already approached above in the section "nanotechnology and environment" due to extreme importance of the subject there considered: the production of electrical energy through nanophotovoltaic panels versus the production through classical means. Obviously this does not exhaust the subject. So, going on, new materials (fullerenes, carbon nanotubes) are now available. No one knows what will be the precise effects when this kind of nanomaterials goes into the environment (and also the effects on health). What are the dangers? Where are they put in the end of their mechanical lives? Will some medical devices be kept inside human body? What are the consequences? Or will be given another destiny to them? Is it possible to predict any kind of effects caused to the environment? Jacobstein (2006) refers that it is important to make an analysis on the risks associated to passive compounds in the less than 100 nanometer size range. There is the possibility, for example, of being introduced inadvertently in human bodies. There is in fact the concern with their ability to be inhaled, absorbed through the skin, or to pass through biological compartment barriers such as the blood brain barrier. This kind of dangers pose a real range of potential health and environmental risks that are associated to their potential toxicity or mutagenicity in their interactions with biological systems (note the analogy of this situation with the considered above in "ethics and nanotecnology in the workplace"). While the range of effects vary, most of the risks may be addressed by advanced industrial hygiene and environmental health practices and techniques that seek to characterize the specific risks, exposure patterns, and control methods and enforce them through a combination of practitioner education, industry self-regulation, monitoring and government regulation. This is an important emerging field in the environmental and health sciences, since most of the existing legislation on environmental, safety, and health risks may cover particulates, but do not take the change in physical and biological properties at the nanoscale into account. It is reasonable to assume that passive nanoscale particle risks, although potentially serious if not addressed, will be characterized and addressed systematically under new versions or extensions to existing occupational, industrial hygiene, environmental, and medical regulations.

## Metaphysical

The incorporation of implants in human body, this is, the incorporation of artificial materials or machines into human systems may conduct to a concern around the human values and principles; will a human being be more than a human being? The borders are unknown. What kind of being and what kind of consequences. How this process may be controlled and accessed? One of the main issues is the "autonomy" of nano-bots, that is: robots whose components are at or near the scale of a nanometre ($10^{-9}$meters), if they are able or not to move and to take decisions autonomously. Are they able to communicate each other and with the external environment? How can human beings ensure to "retire" them at the end of their "missions"? This is a very demanding question. The modification made on human bodies and living systems may create new realities. This is already a very real concern on the society, being in general the society often very skeptic about developments in this area. Catherine Larrère, see citation in Bensaude-Vincent (2010), discusses "the recent trends in nanoethics which anchor ethics in metaphysics or theology by emphasizing the emergence of new relations of men to nature and to God. It is defended that the moral issue raised by the project of enhancing human performances does not really lie in going beyond the boundary of human knowledge and condition. It is more a question of the moral choice underlying this new form of hubris".

Now in another although similar recording, nano-bots, or in general the "convergence" of bio-nano-info-neuro technologies, requires a new contract between science and society; society should not be seen as an afterthought only for making technology assessments, for social acceptability, etc. However, they should be really involved in the process; "converging technologies" - like syn-bio - should not be developed "inside the walls" of research labs only by highly specialized scientists, such challenges should be introduced requiring a cross-disciplinary approach, where project teams include anthropologists, sociologist, philosophers, and ethicists ... with a "precautionary principle" in mind...

Researchers in ethicbots project in EU, which was finished in 2008, worked in order to identify techno-ethical case-studies on the basis of a state-of-the-art survey in emerging technologies for the integration of human and artificial entities; to identify and analyze techno-ethical issues concerned with the integration of human beings and artificial entities, by means of case-studies analysis; to establish a techno-ethically aware community of researchers, by promoting workshops, dissemination, training activities, as well as by the construction of an internet knowledge-base, on the subject of techno-ethical issues emerging from current investigations on the interaction between biological and artificial (software/hardware) entities; to generate inputs to EU for techno-ethical monitoring, warning, and opinion generation (see Capurro et al., 2005-2008).

Rodotà & Capurro (2005) show the limitations on ICT implants in the human body as deriving from an analysis of the principles contained in various legal instruments. They say that they should be assessed further by having regard to general principles and rules concerning the autonomy of individuals, which takes the shape of freedom to choose how to use one's body, *"I am the ruler of my own body"*, freedom of choice as regards one's health, freedom from external controls and influence.

## CONCLUSION

Ethical, legal and social implications in this context show the importance of nanotechnology to the society and the consequences that the development of nanotechnology possibly will bring to mankind in the future. The irreversibility and long-term impact of these new developments enforce to have in mind that it is necessary to take into account also the rights of future generations and the planet (like Hans Jonas' "Principle of responsibility").

Indeed, the future is there and came quick and loudly. Even before the right time. The innovations are going fast and sprout from everywhere. And although the endemic crisis, mainly economic but also social and politic, will bring some restrictions to the investments in health and research areas in many countries, new developments in this area will prevail and many discoveries will be made. The machine is very well assembled and perfectly lubricated: innovations are growing and development is there, as much as the will to go faster and further in many domains of scientific investigation. The research on this subject gets new results every day and new challenges have to be faced. Some work is going on this way and it is intended, in a forthcoming occasion, to analyze some data in order to detect and study the impact of the actual crisis on nanotechnology and mainly on applications to medicine and industry. The crisis may bring some restrictions, but there is a strong will to face the challenges and a new beginning is there.

Ethics concerning nanodevices and the borders of human beings considering nanotechnology in general and applied to medicine (where these concerns assume usuallya more dramatical aspect) also need a new research to be developed next. This debate of the borders of the human being regarding the application of nanotechnology may be also enlarged in order to have a study of the impact into telemedicine and e-health.

The ethical issues related with the themes "nanotechnology and environment" and "nanotechnology in the workplace" are extremely critical, difficult to approach subjects whose development is hard to predict, due to the lack of knowledge of the situations. It can only be filled in the future after observing the practical situations. However, it all requires the try to create credible scenarios that can frame future decision-making processes. It is also urgent, in both theses cases and the above referred, to think about what kind of authorities, possibly with a regulatory vocation, will need to be created to supervise the activities that will appear in the field of nanotechnology.